# AN OPEN-SOURCE, THREE-DIMENSIONAL GROWTH MODEL OF THE MANDIBLE

*On behalf of MAGIC Amsterdam*

*M*andibular *A*natomy & *G*rowth *I*nterdisciplinary *C*onsortium Amsterdam

Cornelis Klop[a], Ruud Schreurs[a,b], Guido A De Jong[b], Edwin TM Klinkenberg[a], Valeria Vespasiano[a], Naomi L Rood[a], Valerie G Niehe[c], Vidija Soerdjbalie-Maikoe[d,e], Alexia Van Goethem[d], Bernadette S De Bakker[f], Thomas JJ Maal[a,b], Jitske W Nolte[a], Alfred G Becking[a]

[a]Department of Oral and Maxillofacial Surgery, Amsterdam UMC and Academic Centre for Dentistry Amsterdam (ACTA), University of Amsterdam, Amsterdam Movement Sciences, Meibergdreef 9, 1105 AZ Amsterdam, The Netherlands

[b]Department of Oral and Maxillofacial Surgery 3D Lab, Radboud University Medical Centre Nijmegen, Radboud Institute for Health Sciences, Geert Grooteplein Zuid 10, 6525 GA Nijmegen, The Netherlands

[c]Department of Radiology, Groene Hart Ziekenhuis, Bleulandweg 10, 2803 HH Gouda, The Netherlands

[d]Department of Forensic Medicine and Pathology, Antwerp University Hospital, Drie Eikenstraat 655, 2650 Edegem, Belgium.

[e]Netherlands Forensic Institute, Department of Forensic Medical Research, Laan van Ypenburg 6, 2497 GB The Hague, The Netherlands.

[f]Department of Obstetrics and Gynecology, Amsterdam UMC location University of Amsterdam, Amsterdam Reproduction and Development research institute, Meibergdreef 9, 1105 AZ Amsterdam, The Netherlands

**CORRESPONDING AUTHOR**

Cornelis Klop, Amsterdam UMC (location AMC), room D2-129, Meibergdreef 9, 1105 AZ Amsterdam

c.klop@amsterdamumc.nl

**ABSTRACT**

The available reference data for the mandible and mandibular growth consists primarily of two-dimensional linear or angular measurements. The aim of this study was to create the first open-source, three-dimensional statistical shape model of the mandible that spans the complete growth period. Computed tomography scans of 678 mandibles from children and young adults between 0 and 22 years old were included in the model. The mandibles were segmented using a semi-automatic or automatic (artificial intelligence-based) segmentation method. Point correspondence among the samples was achieved by rigid registration, followed by non-rigid registration of a symmetrical template onto each sample. The registration process was validated with adequate results. Principal component analysis was used to gain insight in the variation within the dataset and to investigate age-related changes and sexual dimorphism. The presented growth model is accessible globally and free-of-charge for scientists, physicians and forensic investigators for any kind of purpose deemed suitable. The versatility of the model opens up new possibilities in the fields of oral and maxillofacial surgery, forensic sciences or biological anthropology. In clinical settings, the model may aid diagnostic decision-making, treatment planning and treatment evaluation.

1. **INTRODUCTION**

Growth and development of the human body has intrigued humanity for thousands of years. Although curiosity might have sparked this interest at first, studying growth and development has now become a vital scientific discipline. The current possibilities of quantifying and documenting measurements of the human body allow us to gather immense amounts of data. When processed appropriately, this data constitutes reference or normative data, which characterizes what is common in a defined population.

Reference data relating to the mandible came into existence in the second half of the 20$^{th}$ century. One of the first to investigate the normal growth of the mandible was the Swedish orthodontist Arne Björk. Thanks to his unorthodox implant studies, improved understanding of normal growth patterns of the mandible was gained [1,2]. The works of Skieller, Ødegaard and Baumrind have also contributed to new insights in the growth of the mandible through the use of two-dimensional imaging techniques [3–6]. Later studies exploited large longitudinal databases of lateral cephalograms (Fels Longitudinal Study, Bolton-Brush Growth Study) and comprised the first reference data for the mandible [7–9]. These studies report on various linear and angular measurements and assess the growth of the mandible in two dimensions.

Two-dimensional data, however, is merely able to partially represent the growth of a complex, three-dimensional anatomical shape. Only three-dimensional reference data may be able to characterize mandibular growth to its full extent. While there are numerous studies that have modelled the three-dimensional shape of the adult mandible, only a handful of papers do so for the growing mandible [10,11,20–29,12–19]. Most of these studies suffer from limitations, such as a small sample size [10,11,28,12–14,19,20,22,24,27], limited age range [10,20,23–29], potential selection bias [11,12,19,22], or the requirement of extensive manual input, such as landmark annotation [10,20–26,29]. Two studies by Chung et al. focus mainly on the technical framework for building a statistical shape model of the growing mandible [15,16]. Two publications by Coquerelle et al. have respectable samples sizes, but probably not enough to consider it as valid reference data for all ages and both sexes [17,18]. Our own research consortium has mapped the shape of the growing mandible before, but this was done using cadaveric remains [30]. Exact age, sex and reason of death were unknown for these specimens. It is therefore questioned to what extent this growth model may be used in clinical practice as reference data.

The aim of this study was to create a statistical shape model of the growing mandible that encompasses the entire growth period. A secondary goal was to make this model available open-source for further research on the growing mandible and to aid forensic and clinical interpretation and decision-making.

## 2. METHODS

### 2.1. Study samples

Study samples were collected from three post-mortem forensic databases and two clinical databases. Computed tomography (CT) scans of children and young adults (0 to 22 years of age) were considered for inclusion. Further inclusion criteria were a field-of-view large enough to capture the entire mandible and a slice thickness of 1.25 mm or lower. Exclusion criteria were mandibular trauma, evident history of mandibular surgery, and congenital or acquired disorders that could affect the mandible. Scans with movement artefacts or otherwise low quality were also excluded. All included scans were transferred in digital imaging and communications in medicine (DICOM) format, and the chronological age and biological sex of the respective subjects were documented. The accuracy with which the age was reported varied from one day to one month, depending on the database and the age of the sample. For eleven subjects, the exact age was unknown. The age and sex distribution of the complete dataset can be found in Figure 1. A total of 678 samples were included with an overall male-to-female ratio of approximately 2:1 (66% male, 34% female). All methods were carried out in accordance with relevant medical-ethical guidelines and regulations. Further information on the imaging databases can be found in Table 1.

This study uses data from the New Mexico Decedent Imaging Database (NMDID), which originates from the United States [31]. In order to better represent the demographics of the Dutch population, only subjects categorized as "white, non-hispanic, non-latino" were included from this database, many of whom have European ancestry [32]. Since there was an abundance of data of infants (0 to 1 year of age) available in other databases with a lower age uncertainty, infants were excluded from the NMDID database. Reflecting the very nature of forensic data, samples with younger ages (0-3 years) and older ages (13-22 years) were overrepresented compared to the intermediate group (3-13 years). The clinical database of the Amsterdam UMC, location AMC (Amsterdam Medical Center) was only examined for additional samples in the age of 3 to 13 years in order to (partially) compensate for this. The clinical database of the Radboudumc was examined for samples in the age of 3 to 18 years for the similar reasons.

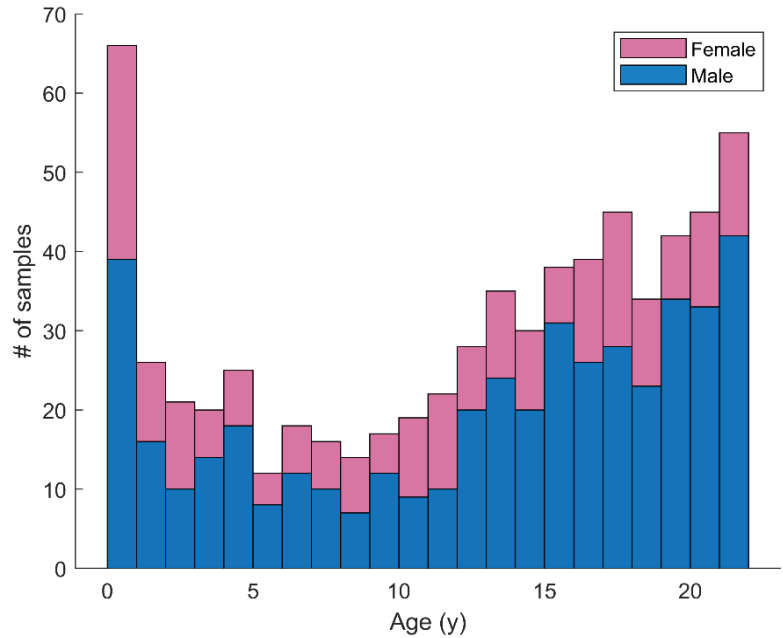

**Figure 1:** Age and sex distribution of the included samples.

### 2.2. Data pre-processing

The mandibles were segmented using either one of two methods: automatic segmentation using the Relu Virtual Patient Creator (Relu BV, Leuven, Belgium) for subjects >1 year of age, or semi-automatic segmentation in Brainlab Elements (Brainlab AG, Munich, Germany) for subjects <1 year of age. The Relu Virtual Patient Creator (creator.relu.eu) allows for fast and automatic segmentation of anatomical models using artificial intelligence (AI)-assisted software. Minor manual corrections were done if deemed necessary. At the time, the Relu software was not trained on data of infants; these scans were therefore segmented using a semi-automatic method in Brainlab Elements. In those cases, one of two observers (CK, NLR) used a combination of thresholding and the *Smart Brush* function to segment the mandible. The mandibular symphysis was included in the segmentation if two separate hemimandibles were still present in younger samples, to produce semantically similar segmentations across the entire age range. Regardless of the segmentation method used, internal structures such as trabecular bone, mandibular canals and tooth buds were filled; only the outer shell of the cortical bone layer was segmented. After segmentation, the mandible was exported as a stereolithographic (STL) 3D model for further processing. Subsequently, the dental area, comprised of erupted teeth and alveolar bone, was manually selected by one of three observers (CK, VV, ETMK) and removed in Meshmixer (Autodesk Inc., San Rafael, CA, USA). This process was done to minimize the impact of the dental region on the growth model.

**Table 1:** Information on imaging databases.
MERC = medical ethics review committee
CME = committee for medical ethics

| Database | No. of samples | Age range (y) | Age accuracy | Type | Country of origin | Max. slice thickness (mm) | Scan period | Medical ethics review |
|---|---|---|---|---|---|---|---|---|
| Netherlands Forensic Institute (**NFI**) | 148 | 0-20 | Samples <1y: one week  Samples>1y: one month | Post-mortem forensic | The Netherlands | 1.0 | 2007 - 2020 | Amsterdam UMC MERC file number 21.477 |
| New Mexico Decedent Imaging Database (**NMDID**) | 318 | 1-22 | All samples: one month | Post-mortem forensic | United States (white, non-hispanic, non-latino) | 1.0 | 2010 - 2017 | See nmdid.unm.edu/resources/data-information |
| Antwerp University Hospital (**UZA**) | 23 | 0-22 | Samples <½y: one day  Samples >½y: one month | Post-mortem forensic | Belgium | 1.25 | 2020 - 2022 | Antwerp University Hospital CME file number 5762 |
| Amsterdam University Medical Center, location AMC (**AMC**) | 110 | 3-13 | All samples: one day | Clinical | The Netherlands | 1.0 | 2012 - 2022 | Amsterdam UMC MERC file number 22.260 |
| Radboud University Medical Center (**Radboudumc**) | 79 | 3-18 | All samples: one day | Clinical | The Netherlands | 1.0 | 2013 - 2022 | Radboudumc MERC file number 2023-16605 |

### 2.3. Statistical shape model

In order to achieve point correspondence among all samples, a symmetrical template with 20,863 points was mapped onto each sample. This template was based on the average mandible of a previous study on the shape of the growing mandible [30]. All samples were roughly pre-aligned to a standardized baseline position in order to improve the results of the template-to-target registration process. The template-to-target registration was done in MATLAB (version 2019a, MathWorks, Natick, MA, USA) and comprised the following steps: rigid registration, followed by non-rigid registration using the MeshMonk algorithm [33]. Rigid registration was done by scaling the template to match the bounding box dimensions of the target. The MeshMonk algorithm involves various smoothing/regularization parameters, which control the coherence of point movement during registration. A number of preliminary experiments were performed to find the optimal values for these parameters. The result of the template-to-target registration was a description of all target meshes with the 20,863 points of the reference template.

After template-to-target registration, the 678 mappings were rigidly aligned onto the template using the Procrustes algorithm. The registration was done using two different methods: 1) with translation and rotation only, thereby retaining the scaling differences between the samples, and 2) with translation, rotation and uniform scaling, thereby removing the scaling differences between the samples. Each mapping was converted into a shape vector of size 62,589 × 1 (20,863 points multiplied by 3 dimensions), regarding each coordinate as an independent variable. Concatenating all shape vectors yielded the shape matrix of size 62,589 × 678. In accordance with the previous step, two shape matrices were constructed; one with original scaling of the models ($SM_{original}$) and one with rescaled models ($SM_{rescaled}$). The shape matrices are openly accessible on the following repository: zenodo.org/doi/10.5281/zenodo.8340160.

### 2.4. Principal component analysis

The shape matrices were subjected to principal component analysis (PCA) to summarize and visualize the variation across the dataset. The PCA on $SM_{original}$ was done to gain insight in the overall growth of the mandible. However, retaining the original scaling of the mandibles would also mean that the variation in larger (hence older) mandibles would weigh heavier in the analysis. This model would therefore mainly reflect the variation in older mandibles. In order to have a fair investigation in the variation across the entire dataset (e.g., male-female differences), PCA was also done on $SM_{rescaled}$. This second model represents variation in younger samples as much as variation in older samples. The PCA results are also published open-access on the aforementioned repository: zenodo.org/doi/10.5281/zenodo.8340160.

### 2.5. Workflow validation

Automatic segmentation of the mandible using the Relu Virtual Patient Creator has been validated in an earlier publication [34]. The inter-observer variability of the semi-automatic segmentation process was validated by calculating the Dice coefficient after segmentation of ten randomly selected mandibles [35]. The removal of erupted teeth and alveolar bone in the 3D models of the mandibles was validated in an earlier study [30].

The template-to-target registration process was validated using two methods: 1) geometrical validation and 2) anatomical validation. Geometrical validation was done by quantifying the distance between the mapping and the original 3D model for 40 randomly selected models. For each point on the mapping, the Euclidean distance to the nearest point on the surface of the original model was documented and averaged over all points. This value indicates how accurate the mapping reproduces the physical shape of the original model. Since an adequate result on the geometrical validation would not guarantee that anatomical landmarks are assigned to their expected locations, anatomical validation was performed as well. One observer (ETMK) identified twelve recognizable anatomical landmarks (Table 2) on 40 randomly selected mandibles. Two observers (CK, ETMK) reached consensus as to which points on the template represented these anatomical landmarks. The Euclidean distance was calculated between the MeshMonk-designated landmarks and the observer-designated landmarks. This value indicates how accurate the MeshMonk-mapping assigns anatomical landmarks to their expected locations. Manual landmark annotation was repeated by one observer (CK) for ten mandibles in order to assess the inter-observer variability of this process.

### 2.6. Statistical analysis

Relationships between principal components (PCs) and age were analyzed in MATLAB using Spearman's rank correlation test ($r_s$). A Student's t-test was used to investigate significant male-female differences in PCs. Because of the large number of t-tests in this study, the standard significance threshold of 0.05 was adjusted with the Holm-Bonferroni method to counteract the overall probability of type I errors and protect against type 2 errors [36].

Table 2: Landmarks used for anatomical validation of the MeshMonk algorithm.

| Landmark | Description |
| --- | --- |
| Pogonion | Most forward-projecting point on the mandibular symphysis |
| Menton | Most inferior point on the mandibular symphysis |
| Gonion right | Midpoint of the mandibular angle |
| Gonion left | |
| Coronoid right | Most superior point on the coronoid process |
| Coronoid left | |
| Sigmoid notch right | Most inferior point on the sigmoid notch |
| Sigmoid notch left | |
| Condyle lateral right | Most laterally-projecting point on the condylar process |
| Condyle lateral left | |
| Condyle medial right | Most medially-projecting point on the condylar process |
| Condyle medial left | |

## 3. RESULTS

### 3.1. Workflow validation

The inter-observer variability of the semi-automatic segmentation process between two observers returned an average Dice coefficient of 0.97 (range of 0.94-0.98). The geometrical validation of the template-to-target registration process demonstrated an average distance between the MeshMonk mapping and the original model of 0.05 mm (99% < 0.25 mm). The anatomical validation showed that the mean distance between the MeshMonk-assigned landmarks and observer-assigned landmarks was 1.1 mm (86% < 2.0 mm). The pogonion and left and right gonion demonstrated the largest mean discrepancy (1.3 mm, 1.7 mm, and 1.8 mm, respectively). Correspondingly, these were also the three landmarks with the largest mean distance between two observers (2.4 mm, 3.1 mm, and 4.6 mm, respectively). On average, there was 1.5 mm discrepancy (76% < 2.0 mm) in manual landmark designation between two observers.

### 3.2. Principal component analysis

PCA on the $SM_{original}$ yielded a very compact model; the first PC alone represented 92% of the total variation and combining the first three PCs explained 95% of the variation. In the PCA model on $SM_{rescaled}$, the first 24 PCs represented 95% of the total variation in the dataset. Compactness plots of both models are shown in Figure 2.

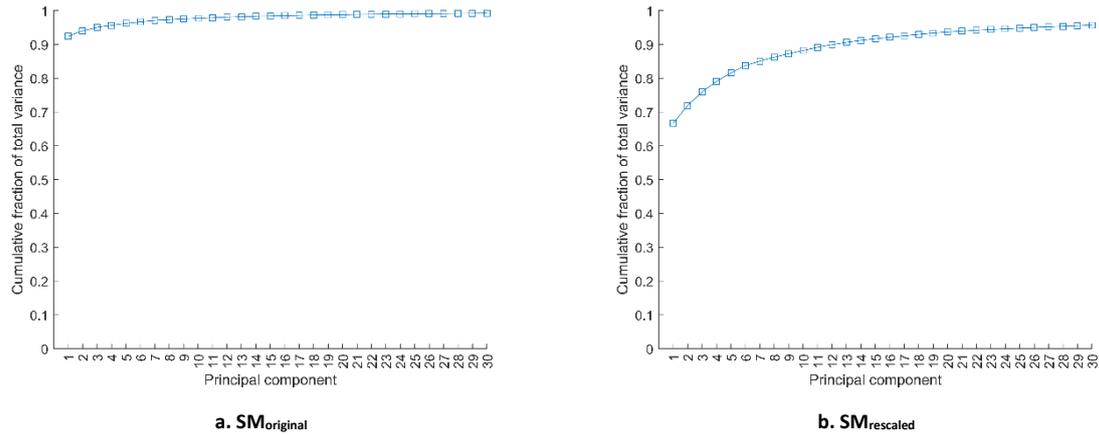

**Figure 2:** Compactness plot of SM$_{original}$ (**a**) and SM$_{rescaled}$ (**b**).

### 3.2.1. Age-related changes

The first PC of SM$_{original}$ was strongly correlated to age ($r_s$ = 0.87, p < 0.0001, Figure 3a). The most noticeable age-related change was the increase in size. Upon closer inspection, additional age-related processes could be identified. There was an evident decrease in gonial angle, and the chin protruded increasingly forward. The two mental tubercles were indistinguishable at younger age, but these became increasingly recognizable. In frontal view, the shape of the mandible transformed from a triangular shape to a rectangular shape due to the development of gonial eversion. In superior view, the width-to-length ratio of the mandible changed from being almost 2:1 at birth to approximately 1:1 upon reaching adulthood.

The first PC of SM$_{rescaled}$ was strongly correlated to age as well ($r_s$ = 0.82, p < 0.0001, Figure 3b). Since size differences were absent in this model, the other age-related changes were even more evident.

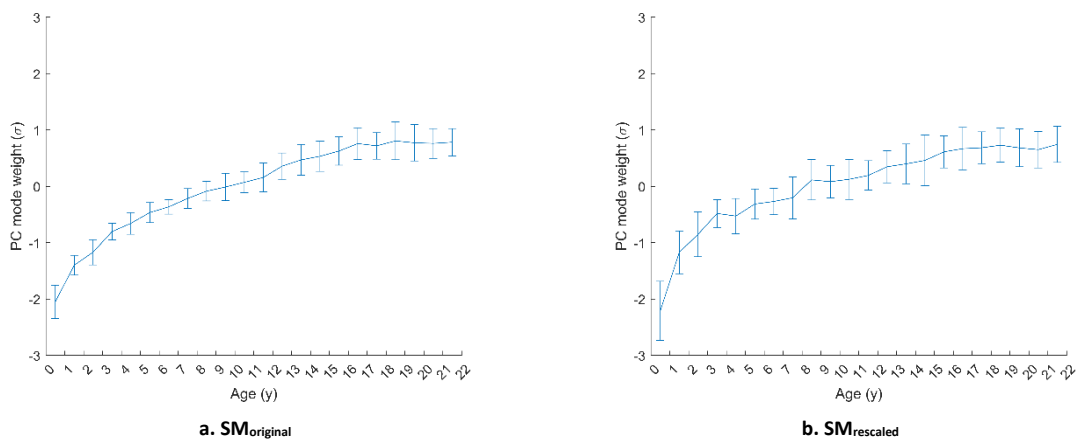

**Figure 3:** Principal component 1 of SM$_{original}$ plotted against age (**a**), principal component 1 of SM$_{rescaled}$ plotted against age (**b**).

### 3.2.2. Sexual dimorphism

Sex-specific variations were studied through PCA on $SM_{rescaled}$. Out of the 677 PCs, four were found to have a significant difference between males and females after Holm-Bonferroni correction. These were PC9 ($p < 0.0001$), PC13 ($p < 0.0001$), PC19 ($p < 0.0001$) and PC22 ($p < 0.0001$). In order to demonstrate the most evident male-female differences, two synthetic models were created in which these four PCs were simultaneously set to "feminine" or "masculine" values (either +3 or -3 standard deviations). All other PCs were kept at a value of 0. Hence, the first model is an example of a mandible with extreme feminine traits, whereas the second model represents a mandible with extreme masculine traits. Overlays of both models are shown in Figure 4 and a description of the differences is provided in the subscript. The average male and female mandible for various ages is shown in Figure 5.

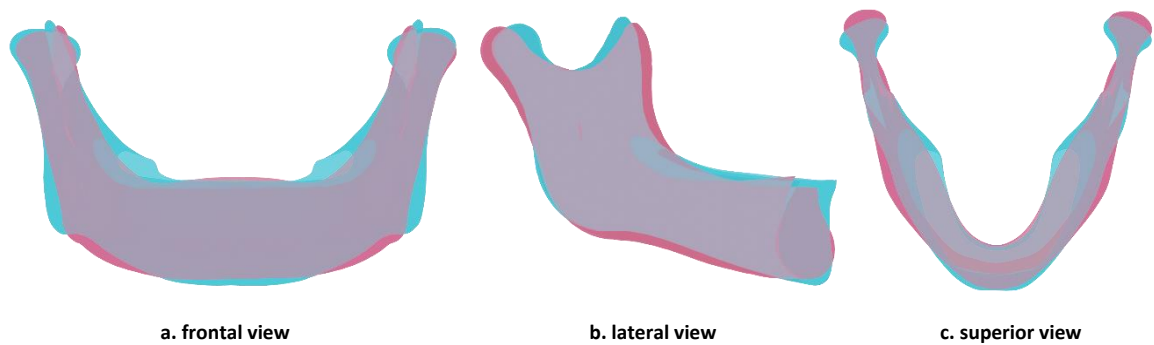

**a. frontal view**  **b. lateral view**  **c. superior view**

**Figure 4:** Overlay of two models with either extreme feminine traits (pink) or masculine traits (blue). These models represent the combined effect of the four principal components (PC9, 13, 19 and 22) with significant male-female differences. Overlap between the two models is shown in pinkish grey. **(a)** Males tend to have a larger gonial eversion, resulting in a rectangular-shaped mandible in frontal view, while females gravitate towards a triangular shape. **(b)** Symphyseal height is generally larger in males than in females, and the masculine mandible tends to have a deeper antegonial notch. **(c)** The male chin tends to have an angular appearance with two distinct mental tubercles, while a curved, rounded chin and lower mandibular border appears to be a feminine trait. The width of the mandible in the molar region is generally larger in females, which contributes to the rounded shape of the feminine mandible in superior view.

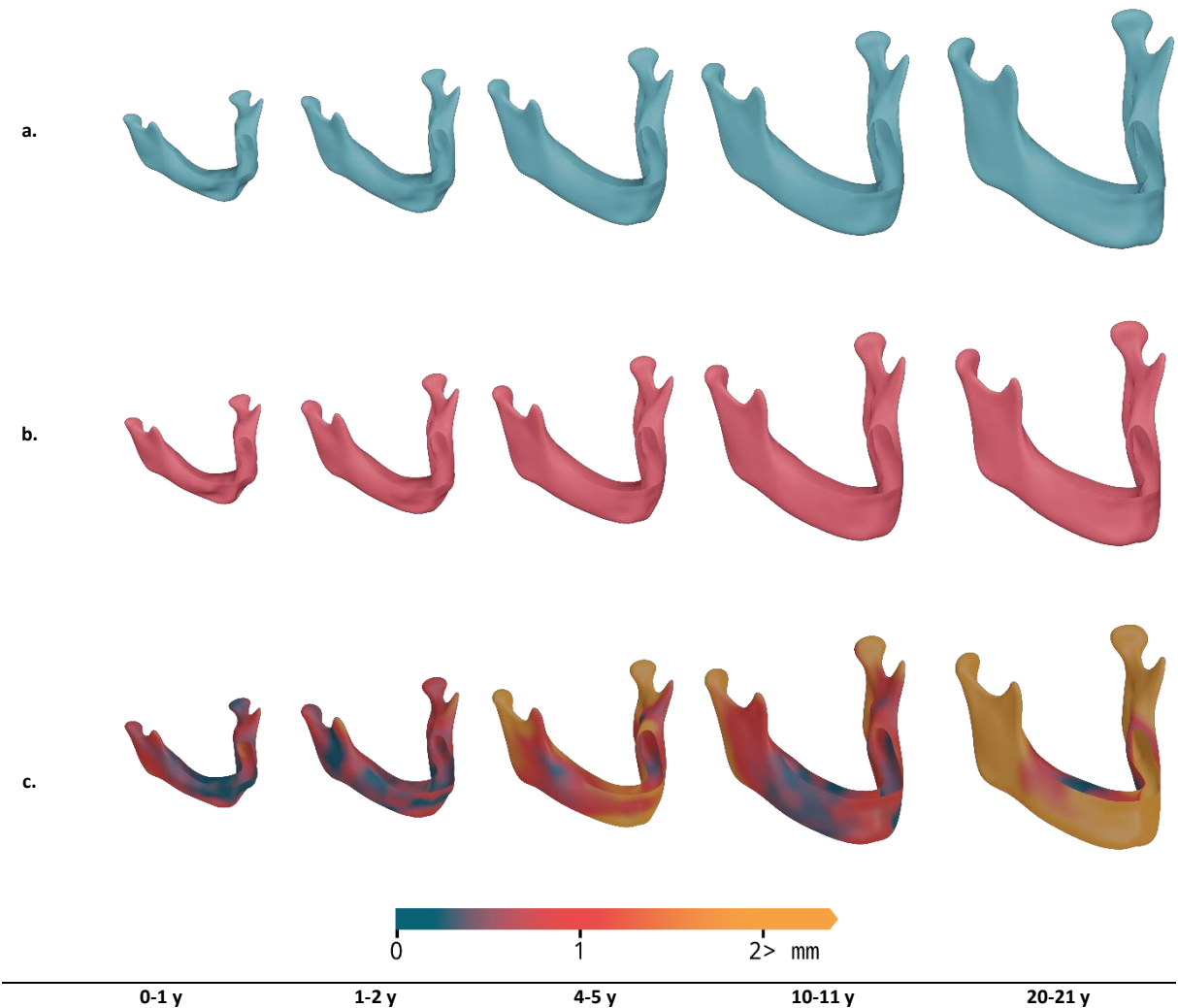

**Figure 5:** The average male **(a)** and female **(b)** mandible at different ages in three-quarter view. A distance map **(c)** between the respective models gives insight in the differences between the male and female mandible at different ages.

## 4. DISCUSSION

In the current study, the first three-dimensional, open-source growth model of the mandible that spans the complete growth period is presented. The validation studies indicated that the utilized methods are robust, accurate and reliable. The geometrical validation yielded excellent results that can barely be improved upon. The anatomical validation showed some discrepancy between the MeshMonk-assigned pogonion and gonion landmarks compared to the manually-assigned landmarks. This is most likely due to the observer's ineptitude to assign these landmarks, as was indicated by the high inter-observer variability of these landmarks. Considering this, the results of the anatomical validation are well within acceptable limits.

Unsurprisingly, age-related changes were the predominant factor of variation in the PCA models on both $SM_{original}$ and $SM_{rescaled}$. The significant relation between age and the first PC in the rescaled model (SM$_{rescaled}$) confirmed that growth encompasses much more than merely an increase in size. Most of the phenomena observed in this study have been reported in literature, such as decrease in gonial angle [21,23,37,38], increase in chin protrusion [17,23,39], increase in gonial eversion [37], and development of mental tubercles [17]. The fact that these findings could also be found in the current study, endorses the validity of the presented model.

The most apparent male-female differences were included in a minority of the PCs. In general, the male mandible was larger than the female mandible, a finding that is supported by other publications [21,23]. The model strongly suggested that gonial eversion is a masculine trait, but this has been debated in literature [40–45]. In the present study, the feminine mandible generally had a shallower antegonial notch and a smaller symphyseal height than the masculine mandible. Both of these observations are corroborated by literature [17,21,23,44–46]. It has been known that the masculine chin tends to have an angular appearance with two distinct mental tubercles, while the feminine chin generally has a curved, rounded shape [43,45,47], a finding that is supported by the present study.

Although many of the age-related and sex-specific variations found in this study have been described in literature before, an important addition of this study is that it is now possible to quantify these variations, relate the changes to reference data, and visualize changes in all three dimensions. The presented growth model has implications in different fields. First, the model can be used in biological anthropology to study ancient populations. Comparing the contemporary shape and development of the mandible to those of ancient populations may reveal changes in morphology over time. These studies may reveal how humans have evolved and how this relates to food intake, daily habits, nutritional conditions, and overall lifestyle.

The statistical shape model may also be applied in forensic investigations. Main applications can be age estimation and sex determination of skeletal remains after mass disasters, war crimes, or in criminal investigations. According to earlier studies, the pelvis and the mandible are among the bones with the most evident sexual dimorphism [48,49]. In any case where DNA analysis is unusable or otherwise unreliable, a post-mortem CT scan of the subject could be obtained, and the growth model could aid in determining the age and sex of the individual.

The most obvious clinical application of the presented model lies in the field of oral and maxillofacial surgery. The availability of three-dimensional reference data accommodates an earlier and deliberate diagnosis of growth disorders. Patients with a suspected hypoplasia or hyperplasia can be compared to the average age-specific and sex-specific mandible to evaluate the extent and location of the disorder. This is also applicable to asymmetrical growth disorders, as the model may be used for determining the affected area and evaluating the severity of the asymmetry. Follow-up of young asymmetrical patients over time is complicated due to simultaneous growth-related changes. By overlaying models of the patient's mandible with their respective age-specific and sex-specific average, it may become easier to establish if the asymmetry has progressed over time. The statistical shape model can also aid in personalized treatment planning, for example in distraction osteogenesis - the model enables the calculation of the desired growth vector in young patients with hypoplasia in order to normalize the shape of the mandible. Another application in treatment planning may include orthognathic surgery, as the model may enable personalized treatment planning towards a symmetrical and harmonious mandible. Additionally, the model may be applied in reconstruction cases after (bilateral) trauma, ablation or radiation therapy, or in oncologic conditions, where traditional mirroring techniques are hampered by the lack of a non-affected side. By making the model publicly available, free of charge, the authors intend to facilitate scientific progress by other groups in any of the aforementioned fields.

In this study, only the chronological age and biological sex of the subjects were documented. Besides these characteristics, some earlier publications also mention race or ethnicity as an independent variable [40,48,50,51]. Although it has been generally accepted within the scientific community that the existence of biological races is an obsolete concept [52,53], there might be differences in mandibular morphology between groups of different ethnic origin. In the current study, the ethnicity of the subjects was not considered, as there are several inherent issues with it. One problem with this type of categorization is its inaccuracy. Many subjects would have an ethnic origin that is unclear or untraceable. Moreover, almost all ethnicities have intermixed throughout the course of history, especially in a multicultural population such as the Dutch. Together, the five radiological databases used in this study reflect a multicultural, but predominantly white population. Hence, the presented growth model should be valid in comparable European populations and their genealogical descendants, such as Canada, the United States, Australia and New Zealand.

In future studies, we aim to investigate the use of AI to replace any manual processes, such as pre-aligning the mandible, and removing the dental region. AI may also play a role in enhancing the accuracy of age estimation and sex determination for forensic purposes. In the future, the growth model might be augmented by including the maxillary complex and dentition as well, which may improve the coordination and execution of orthodontics and orthognathic surgery. The authors are dedicated to collect additional samples to consolidate the current growth model, especially for the age categories that are currently scarcely populated. Other research groups are invited to add data to this database as well, in order to establish a broad spectrum of reference data. In the meantime, it is the author's intent that the open-source data will be used for a wide variety of applications.

**OPEN-SOURCE ACCESS & DISCLAIMER**

All geometrical data and corresponding metadata (sex and age) are accessible on the following repository: zenodo.org/doi/10.5281/zenodo.8340160. It is mandatory to refer the use of this database and this paper in all posters, podium presentations, and publications. All data may be used free-of-charge for any purpose deemed suitable. MAGIC Amsterdam is not liable or responsible for any adverse events encountered during the use and application of this database. The presented growth model is not approved by the US Food and Drug Administration (FDA) nor by the EU Medical Device Regulation (MDR). All data is covered by the GNU General Public License (GPL) v3.0.

**COMPETING INTERESTS STATEMENT & FUNDING**

Several grants were obtained by the MAGIC research consortium throughout the course of this study. In all cases, the ultimate purpose and motivation behind receiving these grants was to create an open-source mandibular growth model. Grant issuers were not able to review or control study design and data collection, had no part in the analysis and interpretation of the data, nor in writing the report or the decision to submit the article for publication.

The MAGIC research consortium would like to cordially thank the Strasbourg Osteosynthesis Research Group (SORG) for the recognition with the Maxime Champy Research Grant in 2021 (€4,000), the Amsterdam Movement Sciences of the Amsterdam UMC for the Dare-to-Dream Grant in 2022 (€5,000) and Relu BV (Leuven, Belgium) for using their Virtual Patient Creator with 50% discount (estimated €3,000).

**AUTHOR CONTRIBUTIONS**

**Cornelis Klop:** Conceptualization, Methodology, Validation, Formal analysis, Investigation, Data Curation, Writing – Original Draft, Writing – Review & Editing, Visualization, Project administration, Funding acquisition

**Ruud Schreurs:** Conceptualization, Methodology, Writing – Original Draft, Writing – Review & Editing, Visualization, Supervision, Funding acquisition

**Guido A De Jong:** Methodology, Software, Investigation, Data Curation, Writing – Review & Editing, Visualization

**Edwin TM Klinkenberg:** Validation, Formal analysis, Investigation, Data Curation

**Valeria Vespasiano:** Validation, Formal analysis, Investigation, Data Curation

**Naomi L Rood:** Validation, Formal analysis, Investigation, Data Curation

**Valerie G Niehe:** Investigation, Resources, Writing – Review & Editing

**Vidija Soerdjbalie-Maikoe:** Investigation, Resources, Writing – Review & Editing

**Alexia Van Goethem:** Investigation, Resources, Writing – Review & Editing

**Bernadette S De Bakker:** Investigation, Resources, Writing – Review & Editing, Funding acquisition

**Thomas JJ Maal:** Conceptualization, Methodology, Writing – Original Draft, Writing – Review & Editing, Supervision, Funding acquisition

**Jitske W Nolte:** Conceptualization, Methodology, Writing – Original Draft, Writing – Review & Editing, Supervision, Funding acquisition

**Alfred G Becking:** Conceptualization, Methodology, Writing – Original Draft, Writing – Review & Editing, Supervision, Funding acquisition

**REFERENCES**

1.  Björk, A. Variations in the Growth Pattern of the Human Mandible: Longitudinal Radiographic Study by the Implant Method. *J. Dent. Res.* **42**, 400–411 (1963).

2.  Björk, A. Facial growth in man, studied with the AID of metallic implants. *Acta Odontol. Scand.* **13**, 9–34 (1955).

3.  Skieller, V., Björk, A. & Linde-Hansen, T. Prediction of mandibular growth rotation evaluated from a longitudinal implant sample. *Am. J. Orthod.* **86**, 359–370 (1984).

4.  Baumrind, S., Ben-Bassat, Y., Korn, E. L., Bravo, L. A. & Curry, S. Mandibular remodeling measured on cephalograms. 1. Osseous changes relative to superimposition on metallic implants. *Am. J. Orthod. Dentofac. Orthop.* **102**, 134–142 (1992).

5.  Ødegaard, J. Growth of the mandible studied with the aid of metal implant. *Am. J. Orthod.* **57**, 145–157 (1970).

6.  Ødegaard, J. Mandibular rotation studied with the aid of metal implants. *Am. J. Orthod.* **58**, 448–454 (1970).

7.  Buschang, P. H., Tanguay, R., Demirjian, A., LaPalme, L. & Goldstein, H. Modeling longitudinal mandibular growth: Percentiles for gnathion from 6 to 15 years of age in girls. *Am. J. Orthod. Dentofac. Orthop.* **95**, 60–66 (1989).

8.  Liu, Y.-P., Behrents, R. G. & Buschang, P. H. Mandibular growth, remodeling, and maturation during infancy and early childhood. *Angle Orthod.* **80**, 97–105 (2010).

9.  Nahhas, R. W., Valiathan, M. & Sherwood, R. J. Variation in timing, duration, intensity, and direction of adolescent growth in the mandible, maxilla, and cranial base: the Fels longitudinal study. *Anat. Rec. (Hoboken).* **297**, 1195–1207 (2014).

10. Stratemann, S. A., Huang, J. C., Maki, K., Hatcher, D. C. & Miller, A. J. Evaluating the mandible with cone-beam computed tomography. *Am. J. Orthod. Dentofac. Orthop.* **137**, (2010).

11. Andresen, P. R. *et al.* Surface-bounded growth modeling applied to human mandibles. *IEEE Trans. Med. Imaging* **19**, 1053–1063 (2000).

12. Andresen, P. R., Nielsen, M. & Kreiborg, S. 4D Shape-Preserving Modelling of Bone Growth. *MICCAI'98* 710-719 BT-Medical Image Computing and Compute at (1998).

13. Bro-Nielsen, M., Gramkow, C. & Kreiborg, S. Non-rigid image registration using bone growth

model. in *CVRMed-MRCAS'97* (eds. Troccaz, J., Grimson, E. & Mösges, R.) 1–12 (Springer Berlin Heidelberg, 1997).

14. Chuang, Y. J., Doherty, B. M., Adluru, N., Chung, M. K. & Vorperian, H. K. A Novel Registration-Based Semiautomatic Mandible Segmentation Pipeline Using Computed Tomography Images to Study Mandibular Development. *J. Comput. Assist. Tomogr.* **42**, 306–316 (2018).

15. Chung, M. K., Chuang, Y. J. & Vorperian, H. K. Online Statistical Inference for Large-Scale Binary Images. *Med. image Comput. Comput. Interv. MICCAI ... Int. Conf. Med. Image Comput. Comput. Interv.* **10434**, 729–736 (2017).

16. Chung, M. K., Qiu, A., Seo, S. & Vorperian, H. K. Unified heat kernel regression for diffusion, kernel smoothing and wavelets on manifolds and its application to mandible growth modeling in CT images. *Med. Image Anal.* **22**, 63–76 (2015).

17. Coquerelle, M. *et al.* Sexual dimorphism of the human mandible and its association with dental development. *Am. J. Phys. Anthropol.* **145**, 192–202 (2011).

18. Coquerelle, M. *et al.* The association between dental mineralization and mandibular form: a study combining additive conjoint measurement and geometric morphometrics. *J. Anthropol. Sci. = Riv. di Antropol. JASS* **88**, 129–150 (2010).

19. Hilger, K. B., Larsen, R. & Wrobel, M. C. Growth modeling of human mandibles using non-Euclidean metrics. *Med. Image Anal.* **7**, 425–433 (2003).

20. Kaya, O. *et al.* Describing the mandible in patients with craniofacial microsomia based on principal component analysis and thin plate spline video analysis. *Int. J. Oral Maxillofac. Surg.* **48**, 302–308 (2019).

21. Kelly, M. P. *et al.* Characterizing mandibular growth using three-dimensional imaging techniques and anatomic landmarks. *Arch. Oral Biol.* **77**, 27–38 (2017).

22. Krarup, S., Darvann, T. A., Larsen, P., Marsh, J. L. & Kreiborg, S. Three-dimensional analysis of mandibular growth and tooth eruption. *Journal of Anatomy* vol. 207 669–682 at https://doi.org/10.1111/j.1469-7580.2005.00479.x (2005).

23. Remy, F. *et al.* Morphometric characterization of the very young child mandibular growth pattern: What happen before and after the deciduous dentition development? *Am. J. Phys. Anthropol.* **170**, 496–506 (2019).

24. Remy, F. *et al.* Characterization of the perinatal mandible growth pattern: preliminary results.

*Surg. Radiol. Anat.* **40**, 667–679 (2018).

25. Cevidanes, L. H. S. *et al.* Assessment of mandibular growth and response to orthopedic treatment with 3-dimensional magnetic resonance images. *Am. J. Orthod. Dentofac. Orthop. Off. Publ. Am. Assoc. Orthod. its Const. Soc. Am. Board Orthod.* **128**, 16–26 (2005).

26. Cevidanes, L. H. S. *et al.* Comparison of relative mandibular growth vectors with high-resolution 3-dimensional imaging. *Am. J. Orthod. Dentofac. Orthop. Off. Publ. Am. Assoc. Orthod. its Const. Soc. Am. Board Orthod.* **128**, 27–34 (2005).

27. Solem, R. C. *et al.* Congenital and acquired mandibular asymmetry: Mapping growth and remodeling in 3 dimensions. *Am. J. Orthod. Dentofac. Orthop.* **150**, 238–251 (2016).

28. Reynolds, M., Reynolds, M., Adeeb, S. & El-Bialy, T. 3-d volumetric evaluation of human mandibular growth. *Open Biomed. Eng. J.* **5**, 83–89 (2011).

29. O' Sullivan, E. *et al.* Growth patterns and shape development of the paediatric mandible – A 3D statistical model. *Bone Reports* **16**, 101528 (2022).

30. Klop, C. *et al.* A three-dimensional statistical shape model of the growing mandible. *Sci. Rep.* **11**, 18843 (2021).

31. Edgar, H. J. H. *et al.* New Mexico Decedent Image Database. Office of the Medical Investigator, University of New Mexico. (2020) doi:10.25827/5s8c-n515.

32. Lao, O. *et al.* Evaluating self-declared ancestry of US Americans with autosomal, Y-chromosomal and mitochondrial DNA. *Hum. Mutat.* **31**, E1875–E1893 (2010).

33. White, J. D. *et al.* MeshMonk: Open-source large-scale intensive 3D phenotyping. *Sci. Rep.* **9**, 6085 (2019).

34. Verhelst, P. J. *et al.* Layered deep learning for automatic mandibular segmentation in cone-beam computed tomography. *J. Dent.* **114**, 103786 (2021).

35. Dice, L. R. Measures of the Amount of Ecologic Association Between Species. *Ecology* **26**, 297–302 (1945).

36. Holm, S. A Simple Sequentially Rejective Multiple Test Procedure. *Scand. J. Stat.* **6**, 65–70 (1979).

37. Ulusoy, A. T. & Ozkara, E. Radiographic evaluation of the mandible to predict age and sex in subadults. *Acta Odontol. Scand.* **80**, 419–426 (2022).

38. Larrazabal-Moron, C. & Sanchis-Gimeno, J. A. Gonial angle growth patterns according to age and gender. *Ann. Anat. = Anat. Anzeiger Off. organ Anat. Gesellschaft* **215**, 93–96 (2018).

39. Upadhyay, R. B., Upadhyay, J., Agrawal, P. & Rao, N. N. Analysis of gonial angle in relation to age, gender, and dentition status by radiological and anthropometric methods. *J. Forensic Dent. Sci.* **4**, 29–33 (2012).

40. Loth, S. R. & Henneberg, M. Gonial inversion: facial architecture, not sex. (2000).

41. Oettlé, A. C., Pretorius, E. & Steyn, M. Geometric morphometric analysis of the use of mandibular gonial eversion in sex determination. *HOMO* **60**, 29–43 (2009).

42. Kemkes-Grottenthaler, A., Löbig, F. & Stock, F. Mandibular ramus flexure and gonial eversion as morphologic indicators of sex. *Homo* **53**, 97–111 (2002).

43. Schutkowski, H. Sex determination of infant and juvenile skeletons: I. Morphognostic features. *Am. J. Phys. Anthropol.* **90**, 199–205 (1993).

44. Saini, V. Scrutiny of four conventional visual traits of mandible for sex estimation in Indian population. *Austin J. Forensic Sci. Criminol.* **4**, 1070 (2017).

45. Ongkana, N. & Sudwan, P. Morphologic indicators of sex in Thai mandibles. *Chiang Mai Med J* **49**, 28 (2010).

46. Schütz, C., Denes, B. J., Kiliaridis, S. & Antonarakis, G. S. Mandibular antegonial notch depth in postpubertal individuals: A longitudinal cohort study. *Clin. Exp. Dent. Res.* **8**, 923–930 (2022).

47. Loth, S. R. & Henneberg, M. Sexually dimorphic mandibular morphology in the first few years of life. *Am. J. Phys. Anthropol.* **115**, 179–186 (2001).

48. Giles, E. Sex determination by discriminant function analysis of the mandible. *Am. J. Phys. Anthropol.* **22**, 129–135 (1964).

49. Loth, S. R. & Henneberg, M. Mandibular ramus flexure: A new morphologic indicator of sexual dimorphism in the human skeleton. *Am. J. Phys. Anthropol.* **99**, 473–485 (1996).

50. Buck, T. J. & Vidarsdottir, U. S. A proposed method for the identification of race in sub-adult skeletons: a geometric morphometric analysis of mandibular morphology. *J. Forensic Sci.* **49**, JFS2004074 (2004).

51. Farkas, L. G., Katic, M. J. & Forrest, C. R. International anthropometric study of facial morphology in various ethnic groups/races. *J. Craniofac. Surg.* **16**, 615–646 (2005).

52. Templeton, A. R. Biological races in humans. *Stud. Hist. Philos. Sci. Part C Stud. Hist. Philos. Biol. Biomed. Sci.* **44**, 262–271 (2013).

53. Yudell, M., Roberts, D., DeSalle, R. & Tishkoff, S. Taking race out of human genetics. *Science (80-. ).* **351**, 564–565 (2016).